\documentclass[apjl]{emulateapj}
\usepackage{apjfonts}

\shorttitle{New Andromeda Satellite}
\shortauthors{Majewski et al.}

\submitted{March 5, 2007}
\journalinfo{Accepted for publication in Astrophysical Journal Letters}
\revised{September 27, 2007}
\accepted{October 1, 2007}
\begin{document}

\title{Discovery of Andromeda XIV: A Dwarf Spheroidal Dynamical Rogue in
  the Local Group? }

\author{
Steven R. Majewski\altaffilmark{1,2},
Rachael L. Beaton\altaffilmark{1,2},
Richard J. Patterson\altaffilmark{1},
Jasonjot S. Kalirai\altaffilmark{3},
Marla C. Geha\altaffilmark{4},
Ricardo R. Mu\~noz\altaffilmark{1},
Marc S. Seigar\altaffilmark{5,6},
Puragra Guhathakurta\altaffilmark{3},
Karoline M. Gilbert\altaffilmark{3},
R. Michael Rich\altaffilmark{7},
James S. Bullock\altaffilmark{5},
 \&
David B. Reitzel\altaffilmark{7}
}

\altaffiltext{1}{Dept. of Astronomy, University of Virginia, P. O. Box
  400325, Charlottesville, VA 22904-4325
  (rlb9n,srm4n,rjp0i,rrm8f@virginia.edu)}

\altaffiltext{2}{Visiting Astronomer, Kitt Peak National Observatory,
  National Optical Astronomy Observatories}

\altaffiltext{3}{UCO Lick Observatory, 1156 High Street, University of
  California, Santa Cruz, CA 95064 (jkalirai,raja,kgilbert@ucolick.org)}

\altaffiltext{4}{NRC Herzberg Institute of Astrophysics, 5071 West
  Saanich Road, Victoria, British Columbia, Canada V9E 2E7
  (marla.geha@nrc-cnrc.gc.ca)}
  
\altaffiltext{5}{Dept. of Physics and Astronomy, 4129 Frederick Reines
  Hall, University of California, Irvine, CA 92697-4575
  (mseigar,bullock@uci.edu)}

\altaffiltext{6}{Department of Physics \& Astronomy, University of
  Arkansas at Little Rock, 2801 S. University Avenue, Little Rock, AR
  72204}

\altaffiltext{7}{Dept. of Physics \& Astronomy, UCLA, Los Angeles, CA
  90095-1562 (rmr,reitzel@astro.ucla.edu)}

\begin{abstract}
  
  In the course of our survey of the outer halo of the Andromeda Galaxy
  we have discovered a remote, possible satellite of that system at a
  projected 162 kpc ($11\fdg7$) radius.  The fairly elongated
  ($0.31\pm0.09$ ellipticity) dwarf can be fit with a King profile of
  1.07 kpc ($d$/784 kpc) limiting radius, where the satellite distance,
  $d$, is estimated at $\sim630-850$ kpc from the tip of the red giant
  branch.  The newfound galaxy, ``Andromeda XIV'' (``AndXIV''),
  distinguishes itself from other Local Group galaxies by its extreme
  dynamics: Keck/DEIMOS spectroscopy reveals it to have a large
  heliocentric radial velocity ($-481$ km s$^{-1}$), or $-206$ km
  s$^{-1}$ velocity relative to M31.  Even at its {\it projected} radius
  AndXIV already is at the M31 escape velocity based on the latest M31
  mass models.  If AndXIV is bound to M31, then recent models with
  reduced M31 virial masses need revision upward.  If not bound to M31,
  then AndXIV is just now falling into the Local Group for the first
  time and represents a dwarf galaxy that formed and spent almost its
  entire life in isolation.

\end{abstract}

\keywords{galaxies: individual (M31, AndXIV) -- galaxies: kinematics and
  dynamics -- Local Group}

\section{Introduction}

Recent discoveries of more than a dozen Local Group (LG) dwarf galaxies
(e.g., Willman et al.\ 2005a,b; Belokurov et al.\ 2006, 2007; Martin et
al.\ 2006, hereafter ``M06''; Zucker et al.\ 2004, 2006a,b, 2007; Irwin
et al.\ 2007) have greatly expanded their known variety by shape,
luminosity, mass, apparent dark matter (DM) fraction, star formation
history (SFH), and internal and bulk dynamics.  The discoveries provoke
consideration of new and/or more complex models for dwarf galaxy
evolution, for the relation between dwarfs and larger (usually parent)
galaxies, the connection between dwarfs and subhalos seen in
$\Lambda$CDM simulations, and the properties of large, $L$* galaxies
themselves.

Despite the variety, all of the newfound dwarf systems (but one ---
Irwin et al.\ 2007) are apparently of dSph type (metal-poor, no young
stars or detected gas) and are Milky Way (MW) or Andromeda (M31)
satellites.  This continues the observational trend that classical LG
dSphs are associated with the LG spirals whereas dIrrs are less
correlated to them, and supports the idea that dSph galaxy evolution
includes ``harassment'' (Mayer et al.\ 2001a,b, 2006), with pronounced
effects on SFHs and mass loss.  Indeed, several newfound dwarfs (e.g.,
Willman I, Ursa Major II, Canes Venatici, Bo\"otes) show evidence for
tidal disruption (Belokurov et al.\ 2006; Willman et al.\ 2006; Zucker
et al.\ 2006b; Fellhauer et al.\ 2007).

The number of newly found systems, and their inferred, often very high
(e.g., Kleyna et al.\ 2005; Mu\~noz et al.\ 2006a) DM fractions, have
important implications for hierarchical galaxy formation models, if the
satellite galaxies are related to ``subhalos'' seen in $\Lambda$CDM
models of structure formation on galaxy-sized scales.  For example, the
``missing satellites'' gap between predicted and known subhalos (Moore
et al.\ 1999; Klypin et al.\ 1999) may be less pronounced than previously
believed and thus our understanding of galaxy formation on the smallest
scales and at the earliest times may need revising (e.g., Bullock et
al.\ 2000, Benson et al.\ 2002, Taylor et al.\ 2004).

Satellite galaxies also trace the gravitational extent of their parent
halos.  Velocity-distance distributions of satellites have long been
used to constrain parent system masses through the Jeans equation (e.g.,
Hartwick \& Sargent 1978; Frenk \& White 1980; Zaritsky et al.\ 1989;
Sakamoto et al.\ 2003; Evans \& Wilkinson 2000; Evans et al.\ 2000).
Such analyses have recently yielded some surprising results,
particularly that the M31 virial mass may be smaller than previously
thought, more similar to --- or possibly even less than --- that of the
MW (e.g., C\^{o}t\'{e} et al.\ 2000; Evans \& Wilkinson 2000; Evans et
al.\ 2000; Geehan et al.\ 2006; Seigar et al.\ 2007, hereafter S07),
though still in the virial mass range expected for a galaxy of its
luminosity (van den Bosch et al.\ 2007).  This result, however, suggests
a rather lower $M/L$ for M31 than MW, since it is now clear
(Guhathakurta et al.\ 2005; Kalirai et al.\ 2006) that M31's inner,
metal-rich spheroid dominates its outer, metal-poor halo (Ostheimer
2003; Chapman et al.\ 2005; Guhathakurta et al.\ 2005; Kalirai et al.\ 
2006; Ibata et al.\ 2007) to a much larger radius than in the MW, while
the overall M31 luminosity is generally found to be twice the MW's (or
more), consistent with M31's much larger globular cluster population
(e.g., Chandar et al.\ 2004).

In the above applications of the Jean's equation distant satellites
generally place the tightest mass constraints.  Thus new finds of remote
M31 satellites (e.g., M06) are valuable for reassessing M31's mass, once
their velocities are known.  In this context, a newly discovered dwarf
galaxy found during our survey of the outermost parts of M31 challenges
the prevailing ``small mass'' models for M31.  This dwarf, ``Andromeda
XIV (AndXIV)'', is a dynamical rogue with the largest $v^2r$ moment ---
by at least a factor of four --- of any satellite ever suspected of
being a part of the M31 system.

\section{Portrait of a Rogue: Andromeda XIV}

AndXIV was found as an obvious cluster of stars at a 162 kpc projected
distance from M31 in a field imaged during our Kitt Peak 4-m+Mosaic CCD
study of the M31 halo.  The new dwarf lies at $(\alpha,\delta)_{2000}$=
($00^{\rm h}51^{\rm m}35\fs0,+29\arcdeg41\arcmin49\arcsec$).  Our survey
uses the Washington $M,T_2$+$DDO51$ photometric system, which
substantially enhances discrimination of (foreground MW) dwarf stars
from RGB stars by the gravity-sensitivity of the $DDO51$ filter (Fig.\ 
1a; Majewski et al.\ 2000).  The images were obtained on UT 2005 Oct 27
during photometric conditions with $T_2$ seeing of $0.9\arcsec$, and net
integrations of ($900$, $900$, $3\times1800$) sec in ($M,T_2,DDO51$).
Photometric measurements made use of the stand-alone version of DAOPHOT
(Stetson 1987).  Numerous Geisler (1990) standard star observations
enable transformation of the instrumental photometry into the standard
system, accounting for color and airmass terms.

\begin{figure}
\plotone{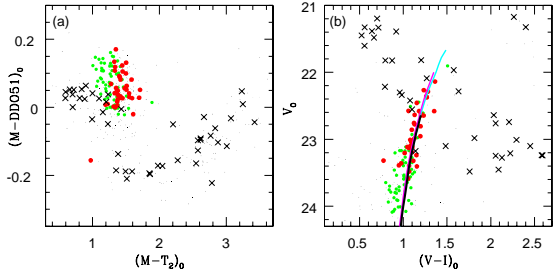}
\caption{(a) Color-color diagram (2CD) and (b) color-magnitude diagram (CMD)
  of a $10\farcm4\times10\farcm4$ CCD field containing AndXIV.  In
  panel (b) the $M,T_2$ photometry is converted to $V,I$ using the
  transformation equations in Majewski et al.\ (2000).  Colored circles
  are all stars selected to be AndXIV giants by joint position in panels
  (a) and (b), with red circles being AndXIV giant candidates so
  selected {\it and} having the AndXIV velocity (plus one extra $v_{\rm
    hel}$ member found among ``slitmask filler'' targets --- the 2CD
  outlier) and green circles other giant candidates not observed
  spectroscopically with Keck+DEIMOS.  The $\times$ symbols correspond
  to other DEIMOS-observed ``slitmask-filler'' stars with non-AndXIV
  velocities.  Small points are other stars in the field.  Several
  fitted Girardi et al.\ (2002) isochrones are included: [Fe/H]=$-1.3$
  at $(m-M)_0=24.3$ (cyan), [Fe/H]=$-1.7$ at $(m-M)_0=24.7$ (magenta),
  and [Fe/H]=$-2.3$ at $(m-M)_0=25.4$ (black).}
\label{f1}
\end{figure}

Similar to previous $M,T_2,DDO51$ studies of MW dSphs (Palma et al.\ 
2003, hereafter P03; Westfall et al.\ 2005; Mu\~noz et al.\ 2005,
2006a,b; Sohn et al.\ 2007) thirty-eight stars selected to be both along
the apparent RGB in the CMD (Fig.\ 1b) and to be $DDO51$-selected RGB
stars (Fig.\ 1a) were subsequently targeted with Keck+DEIMOS multislit
spectroscopy on UT 20-21 Nov 2006.  The spectra have 1.3~\AA\ resolution
(FWHM), are centered at 7800 \AA\ , and include the Ca infrared triplet.
These observations were reduced similarly to the methods described in
Guhathakurta et al.\ (2006) and Simon \& Geha (2007) and are described
in more detail in Geha et al.\ (2007, hereafter G07).  Derived radial
velocities (RVs) from median $S/N=8.4$ per resolution element spectra of
the 38 stars show them to share a common, distinct and extreme
heliocentric velocity, $v_{\rm hel}=-481.1 \pm 1.2$ km s$^{-1}$ (Fig.\ 
2).  That all of the photometrically-selected AndXIV giant candidates
sampled share a common RV demonstrates that stars so-selected likely
trace AndXIV's structure reliably (Fig.\ 3).  Another kinematically
confirmed member is found among 43 other stars randomly observed in the
DEIMOS slitmasks (see Fig.\ 1).  Removing MW motions yields a $-206$ km
s$^{-1}$ AndXIV velocity relative to M31.

\begin{figure}
\plotone{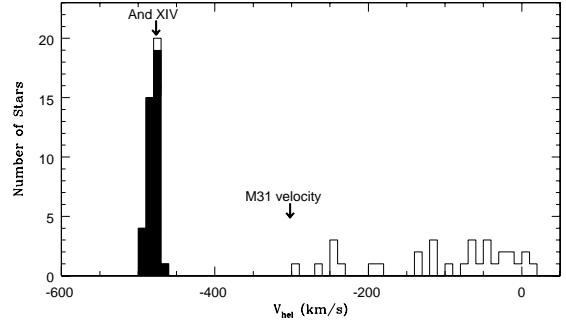}
\caption{The $v_{\rm hel}$ distribution of stars observed with 
  Keck$+$DEIMOS.  The dark histogram is the distribution of
  photometrically-selected AndXIV RGB stars (basically all of the red
  points in Fig.\ 1, except the obvious 2CD outlier), and the clear
  histogram is all other stars with Keck spectroscopy.}
\label{f2}
\end{figure}

AndXIV RGB candidates with and without spectra show an elongated
distribution (Fig.\ 3).  Fitting these with a King profile using the
methods in, e.g., P03 and Westfall et al.\ (2005) yields a
$2.9\pm0.8\arcmin$ core radius, a $4.7\pm0.9\arcmin$ limiting radius,
and $0.31\pm0.09$ ellipticity (Fig.\ 4); the large uncertainties are due
to there being only 101 stars in the top 1.5 mag of the RGB (Fig.\ 1b)
to use in the profile.  The AndXIV structure at large radii is not well
established; whether the apparent ``extra-King profile'' stars mimic
tidal debris features (e.g., Johnston et al.\ 1999; Majewski et al.\ 
2003) or are actually an intrinsic property of the dwarf would give
important clues to the system's interaction history and whether it is on
a bound orbit (see \S 3).

The M31 \ion{H}{1} survey of Braun \& Thilker (2004; e.g., their Fig.\ 
7) shows two compact high velocity clouds near AndXIV, but whose RVs
differ from that of AndXIV by $\sim65$ and $\sim125$ km s$^{-1}$.  The
currently fit structural properties, lack of obvious very young stars
(i.e. a ``blue plume''), and apparently no corresponding \ion{H}{1} all
suggest AndXIV is a dSph galaxy.

From the RGB {\it shape} the AndXIV distance is uncertain due to a
degeneracy of isochrone fits for varying metallicities to what is only
the top 1.5 mag of RGB in Figure 1b.  Based on the $V$ magnitude of the
tip of the RGB, the [Fe/H]=$-1.7$, $(m-M)_0=24.7$ isochrone seems the
best match to the data.  We (G07) are in the process of refining the
AndXIV metallicity estimates using the Keck spectra.  Fortunately,
however, the absolute $I=T_2$ magnitude of the RGB tip varies by only
$M_{T_{2}}=3.80-4.05$ over $-2.3 \le$[Fe/H]$\le -0.3$ (Bellazzini et
al.\ 2004).  The brightest RV member has $T_{\rm 2,o}=20.6$ whereas the
brightest RGB {\it candidate} (Fig.\ 1b) has $T_{\rm 2,o}=20.2$; thus,
the AndXIV distance modulus plausibly ranges from $(m-M)_0=24.0$ (630
kpc) to 24.65 (850 kpc), {\it if} the RGB tip is populated in this low
mass system.  Deeper imaging of AndXIV will clarify its age, metallicity
and distance.

We estimate the AndXIV luminosity by comparison of its luminosity
function in the top 1.5 mag of its RGB to that of other, Galactic dSphs
studied in the same photometric system and using the same techniques.
Comparison to the Carina dSph RGB (Mu\~noz et al.\ 2006b) --- a system
with [Fe/H] $\sim -1.7$ --- yields an AndXIV luminosity of
$1.8\times10^5$ $L_{\sun}$ (for an adopted Carina $L=4.3\times10^5$
$L_{\sun}$).  Comparison to the RGB (P03) of the more metal poor
([Fe/H]$\sim -2.2$) Ursa Minor dSph (with $L=2.9\times10^5$ $L_{\sun}$)
yields an AndXIV luminosity of $2.2\times10^5$ $L_{\sun}$.

\begin{figure}
\plotone{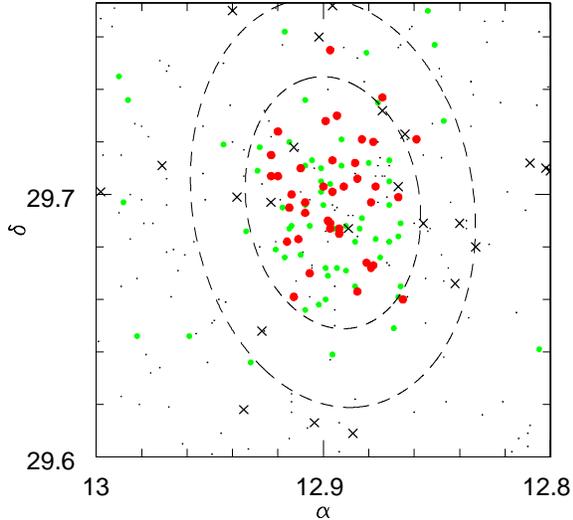}
\caption{Sky distribution (coordinates in degrees) of stars in 
  Fig.\ 1, with same symbols.  Ellipses show core and limiting radii of
  the best-fitting King profile (Fig.\ 4).  The figure's upper edge is
  the edge of our CCD field of view.}
\label{f3}
\end{figure}

\begin{figure}
\plotone{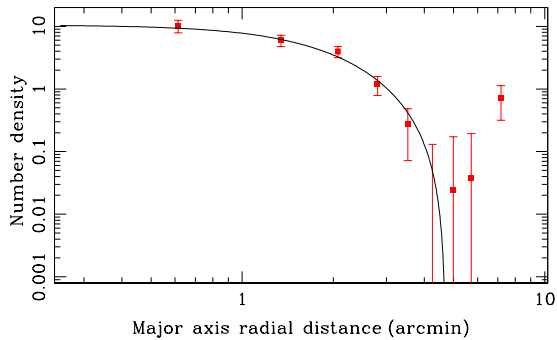}
\caption{King profile (with $r_{\rm c}=2\farcm9$ and $r_{\rm lim}=4\farcm7$)
  fit to colored points in Fig.\ 3 binned by elliptical radius.  The
  density is in stars~arcmin$^{-2}$, after subtraction of a fitted 0.18
  stars arcmin$^{-2}$ background. }
\label{f4}
\end{figure}

\section{A More Massive M31 or a Local Group Interloper?}

Setting a minimum mass on M31 depends crucially on the distance to
AndXIV, if this satellite is bound to its parent. Figure 5 shows the
best-fitting M31 mass model (baryons+DM), represented as the escape
velocity as a function of deprojected (3D) radius from M31's center,
from S07. This analysis uses published H$\alpha$ and \ion{H}{1} rotation
data extending to $\sim$35 kpc to determine a cosmologically-motivated
M31 mass profile. The S07 model also allows for adiabatic contraction in
an initial NFW halo, and is truncated at $\sim$260 kpc, which is the
best estimated virial radius ($R_{\rm vir}$) for M31.  The dashed lines
in Figure 5 represent the $\pm2\sigma$ range for the best-fitting S07
adiabatic contraction models in {\it both} halo concentration ($c=R_{\rm
  vir}/R_{\rm s}$, where $R_{\rm s}$ is the scale length; Bullock et al.
2001) {\it and} virial mass: ($c,M$)=(17.8, $8.0 \times 10^{11}
M_{\sun}$) and (22.2, $8.8 \times 10^{11} M_{\sun}$).  The median
(solid) line corresponds to the best-fitting case with ($c,M$)=(20, $8.4
\times 10^{11} M_{\sun}$).  This median mass is similar to, but slightly
higher than, virial masses based upon satellite kinematics (e.g., $\sim
8 \times 10^{11} M_{\odot}$; C\^{o}t\'{e} et al.\ 2000; Geehan et al.\ 
2006). Both methods yield masses consistent with expectations for a
galaxy with M31 luminosity, where one would expect a range
$(1.4\pm0.7)\times10^{12}M_{\odot}$ based on the 2dF luminosity function
and correlation function (van den Bosch et al.\ 2007).  The virial mass
presented in the $\Lambda$CDM model of Klypin et al.\  (2002) is about
60\% higher than that of S07, i.e. closer to the median of the above
range.  The difference is primarily driven by the fact that S07 use the
3.6-$\mu$m Spitzer image of M31 (Barmby et al.\ 2006) to determine the
baryonic mass distribution, whereas Klypin et al.\ (2002) used optical
light.  One might therefore expect the analysis presented in S07 to be
an improvement over the Klypin et al.\ (2002) model.

\begin{figure}
\plotone{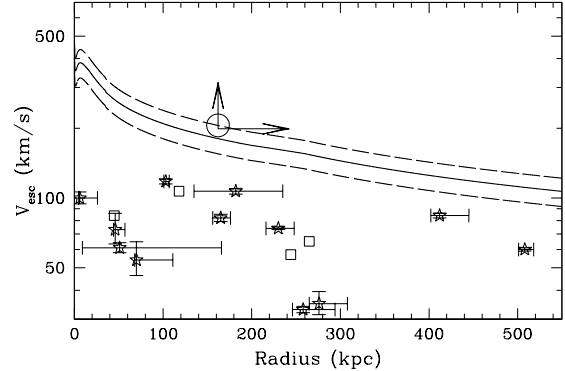}
\caption{Escape velocity versus 3D distance from
  M31 center for the best-fitting M31 mass model of S07.  The star
  symbols represent the positions and reduced velocities relative to M31
  (see text) for commonly accepted Andromeda satellites (e.g., C\^ot\'e
  et al.\ 2000; Evans \& Wilkinson et al.\ 2000); more recent And
  satellite discoveries (Zucker et al.\ 2004, Chapman et al.\ 2005) with
  velocity data are shown as squares.  The circle represents the
  conservative position of AndXIV based on using the {\it projected}
  radius of M31.  If AndXIV is 630 kpc distant (\S 2), its 3-D distance
  from M31 (224 kpc) puts AndXIV $2\sigma$ above the median model escape
  velocity.  }
\label{f5}
\end{figure}

The RVs (reduced to a common projection parallel to the M31-MW vector;
C\^ot\'e et al.\ 2000) of all commonly attributed M31 satellites and
more recent satellite discoveries with relevant data are indicated in
Figure 5.  In most cases, these RVs are less than half escape velocity:
in contrast, the reduced RV lower limit for AndXIV of $-198$ km s$^{-1}$
--- even at AndXIV's {\it projected} 162 kpc M31 distance --- already
places AndXIV beyond the nominal S07 model escape velocity limit.
Because Figure 5 ignores the unknown orthogonal velocity components,
each point only {\it approximates} each satellite's minimum 3-D relative
velocity; in principle, the latter can be overestimated for widely
separated satellites if the {\it M31 system} has a large bulk tangential
motion that projects to a substantial RV along the line of sight to the
satellite.  However, both the ``timing argument'' and arguments based on
M33 water maser proper motions coupled with no apparent M31 tidal
disruption of M33's disk suggest a $<100$ km s$^{-1}$ M31 transverse
velocity, with a northerly motion more strongly ruled out (Loeb et al.\ 
2005); this upper limit (and discordant direction of motion) suggests
little contribution of M31 tangential motion to AndXIV's RV.

On the other hand, considering that two components of AndXIV's
individual velocity are not being accounted for in Figure 5 and that the
3D distance of AndXIV from M31 likely exceeds 162 kpc, one is forced to
conclude either that (1) the prevailing estimates of the M31 virial mass
need upward revision by likely significant amounts, or (2) AndXIV is
presently falling into the M31 potential for the first time.  There is
no significant known galactic mass anywhere along any likely
backward-extrapolated AndXIV trajectory over the last Hubble time if
AndXIV is not bound to M31.

AndXIV lies $\sim3\fdg3$ south of And XIII, the southernmost of
four M31 satellites found by M06.  Based on the coincidences that (1)
these four satellites lie within $2\arcdeg$ of one another despite the
57 deg$^2$ M06 survey area, and (2) the metallicities, sizes and
luminosities of the M06 dSphs are similar, M06 speculate on a possible
linked origin for the group.  Though $\sim4\arcdeg$ from the centroid
of M06 objects, AndXIV does continue to larger radius an alignment of
all five objects along a radial vector from M31's center.  Indeed, all
five objects, {\it as well as} NGC147, NGC185, NGC205, M32, and And I,
lie $<30\arcmin$ ($\lesssim7$ kpc projected distance) from this same
vector across a 19$\arcdeg$ (260 kpc) projected span.  Though still
very uncertain, AndXIV may have a similar distance and metallicity as
the M06 group; however, it is clear that AndXIV is several times larger
and brighter than the M06 dSphs.  Testing a possible association of
AndXIV to the M06 objects, and indeed all of these objects to other
potential ``dynamical families'' suggested for M31 satellites (Koch \&
Grebel 2006; McConnachie \& Irwin 2006; Metz et al.\ 2007), requires
velocities for the M06 objects and more accurate distances to all five
systems.

AndXIV may provide a unique laboratory for LG astrophysics.  If bound to
M31, AndXIV's distance sets a critical lower limit to M31's mass, just
as the inclusion of the distant and high velocity Leo I dSph as a bound
member of the Milky Way strongly affects estimations of its mass (e.g.,
Zaritsky et al.\ 1989; but cf.\ Sakamoto et al.\ 2003).  Geha et al.\ 
(2006) show that external forces seem necessary for complete removal of
gas from dwarf galaxies.  AndXIV's lack of gas and its elongated, low
concentration structure may point to significant environmental (e.g.,
ram pressure, tidal) effects.  The discovery of AndXIV tidal tails would
provide critical additional proof that it is a true M31 satellite.  On
the other hand, if demonstrably unbound to M31, AndXIV would join
Tucana, and possibly Leo I (though see Sohn et al.\ 2007), as an unusual
LG dSph having no prior association with a large spiral and demonstrate
that dSphs can form in isolation without harrassment (\S 1).  In this
case, AndXIV offers the opportunity to explore the internal dynamics,
star formation history and chemical enrichment of a dwarf galaxy before
first interaction.

\acknowledgements

We gratefully acknowledge NSF grants AST 03-07842 and AST 03-07851
(R.L.B., S.R.M., R.J.P., R.R.M.), AST 03-07966 and AST 05-07483 (J.S.K.,
P.G., K.M.G.), AST 03-07931 (R.M.R., D.B.R.), and AST-0607377 (M.S.S,
J.B.).  R.L.B., S.R.M., R.J.P., R.R.M. also appreciate support from
Frank Levinson through the Celerity Foundation.  R.L.B. is grateful for
undergraduate awards from Virginia Space Grant Consortium and University
of Virginia's Harrison Institute.  J. S. K. is supported by Hubble
Fellowship grant HF-01185.01-A, awarded by the Space Telescope Science
Institute, which is operated by the Association of Universities for
Research in Astronomy, Inc., under NASA contract NAS5-26555.  K.M.G.
acknowledges an NSF Graduate Fellowship.  M.S.S.  acknowledges partial
support from a Gary McCue Fellowship through the Center of Cosmology, UC
Irvine.

\end{document}